\documentclass{aa}
\usepackage{txfonts}
\usepackage{graphicx}
\usepackage{natbib}
\bibpunct{(}{)}{;}{a}{}{,}

\begin{document}

\title{Physical limits to the validity of synthesis models}
\subtitle{The Lowest Luminosity Limit.}

\author{M. Cervi\~no \inst{1,2} \and
        V. Luridiana\inst{1}}

\institute{Instituto de Astrof\'\i sica de Andaluc\'\i a (CSIC), Camino bajo
        de Hu\'etor 24, Apdo. 3004, Granada 18080, Spain\\
        \and
        Laboratorio de Astrof\'\i sica Espacial y F\'\i sica Fundamental
        (INTA), Apdo.  50727, Madrid 28080, Spain\\
}

\date{Received 12 February 2003; accepted 29 August 2003}
\offprints{M. Cervi\~no; \email{mcs@laeff.esa.es}}

\abstract{ In this paper we establish a necessary condition for the
application of stellar population synthesis models to observed star
clusters.  Such a condition is expressed by the requirement that the total
luminosity of the cluster modeled be larger than the contribution of the
most luminous star included in the assumed isochrones, which is referred to
as the Lowest Luminosity Limit (LLL).  This limit is independent of the
assumptions on the IMF and almost independent of the star formation
history.  We have obtained the Lowest Luminosity Limit for a wide range of
ages (5 Myr to 20 Gyr) and metallicities ($Z$=0 to $Z$=0.019) from the
\cite{Gi02} isochrones.  Using the results of evolutionary synthesis
models, we have also obtained the minimal cluster mass associated with the
LLL, ${\cal M}^{min}$, which is the mass value below which the observed
colors are severely biased with respect to the predictions of synthesis
models.  We explore the relationship between ${\cal M}^{min}$ and the
statistical properties of clusters, showing that the magnitudes of clusters
with mass equal to ${\cal M}^{min}$ have a relative dispersion of 32\% at
least (i.e., 0.35 mag) in all the photometric bands considered;
analogously, the magnitudes of clusters with mass larger than $ 10 \times
{\cal M}^{min}$ have a relative dispersion of 10\% at least.  The
dispersion is comparatively larger in the near infrared bands: in
particular, ${\cal M}^{min}$ takes values between 10$^4$ and 10$^5$
M$_\odot$ for the $K$ band, implying that severe sampling effects may
affect the infrared emission of many observed stellar clusters.  As an
example of an application to observations, we show that in surveys that
reach the Lowest Luminosity Limit the color distributions will be skewed
toward the color with the smallest number of effective sources, which is
usually the red, and that the skewness is a signature of the cluster mass
distribution in the survey.  We also apply our results to a sample of
Globular Clusters, showing that they seem to be affected by sampling
effects, a circumstance that could explain, at least partially, the bias of
the observed colors with respect to the predictions of synthesis models.
Finally, we extensively discuss the advantages and the drawbacks of our
method: it is, on the one hand, a very simple criterion for the detection
of severe sampling problems that bypasses the need for sophisticated
statistical tools; on the other hand, it is not very sensitive, and it does
not identify all the objects in which sampling effects are important and a
statistical analysis is required. As such, it defines a condition necessary
but not sufficient for the application of synthesis models to observed
clusters.

}

\titlerunning{The Lowest Luminosity Limit}
\maketitle

\section{Introduction and motivation}

The comparison between theoretical models and observations is the basic
procedure that allows the evaluation of our current knowledge of Nature;
underlying this statement is the assumption that better observational data,
by setting more stringent constraints, make such comparison more
meaningful.  Although this is almost always the case, there is a situation
in which, paradoxically, the opposite is true, and the more the data
quality improves, the more biased the theoretical inferences turn out to
be. This is indeed the case of the analysis of the integrated light of
stellar populations. The progressively higher sensitivity of modern
instruments provides access to data from increasingly fainter stellar
clusters, up to a point where we must begin to take into account the
limitations posed by the discreteness of the number of stars in a system:
due to incomplete sampling, both small clusters and small samples of large
clusters have stellar mass distributions that may differ substantially from
the one predicted by the underlying initial mass function (IMF).
Nevertheless, most theoretical models assume that the IMF of clusters is
completely populated, i.e., that the distribution of stellar masses is
continuous and that all the evolutionary stages are well sampled.
Obviously, any model assuming a continuous IMF (hereinafter, analytical
models) will be correct only under the asymptotic assumption of an infinite
number of stars. Hence, the validity of the comparison between the
predictions of synthesis models and real systems, where the IMF is not
perfectly sampled, depends on the size of the system.

Several works have been written that deal directly or indirectly with the
subject of sampling effects.  For example, \cite{BB77}, \cite{CBB88},
\cite{GB93}, and \cite{GCBB95}, who show the relevance of sampling effects
for the study of LMC clusters; \cite{SF97}, who determine how sampling
effects affect integrated near-infrared colors; \cite{CLC00} and
\cite{Cetal00}, who study the effects of sampling in some observables of
young star-forming regions; \cite{Cetal01} and \cite{CM02}, who estimate
the effects of sampling in stellar yields and chemical evolutionary models.
Sampling effects also underlie the study of surface brightness fluctuations
\citep{TS88}, a primary distance indicator that is based on the analysis of
the variations with distance of the amount of stars sampled by CCD pixels:
for example, \cite{enzo} show that surface brightness fluctuations may
suffer from a bias that depends on the density of stars in the image
pixels.

In almost all the preceding works, sampling effects have been evaluated by
the use of Monte Carlo simulations. Alternatively, \cite{CVGLMH02} proposed
a formalism, based on the original formulation by \cite{Buzz89}, where the
mean value, the dispersion, and the correlation coefficient of different
observables are obtained analytically using a continuously distributed IMF.
The method is applied to young star-forming regions ($t < 20$ Myr) and
compared with the results of Monte Carlo simulations, showing that in both
cases the results are quite similar, except for colors and equivalent
widths in clusters with a low number of stars.  The method is extensively
tested in \cite{CVG02} for clusters with a number of stars between 1 and
10$^3$, where it is shown that it reproduces the average value and the
dispersion of quantities obtained from Monte Carlo simulations, i.e., the
luminosities of Monte Carlo simulations are {\it distributed around} the
mean value obtained from the analytical model, {\it if the quantity scales
linearly with the amount of stars in the system}.  However, when the
modeled properties are logarithmic quantities or ratios, as in the case of
equivalent widths and colors, the mean values of Monte Carlo simulations
are {\it biased} with respect to the results of the analytical modeling;
the smaller the system, the more severe the bias. Unfortunately the authors
are not able to quantify this bias in an analytical way for very small
systems.

The subject of sampling has also been addressed by, e.g., \cite{LM00},
\cite{Gproc00}, \cite{Cgal1}, \cite{Brutuc}, \cite{Gituc}, \cite{Cmpe},
\cite{Cpeim,Cgal2}, \cite{CVGcoim}, and \cite{Cgtc,Cgal3,Cter,Clan}.
However, since all these works are published in conference proceedings,
their consequences are not extensively explored.  \cite{LM00} evaluate
sampling effects on monochromatic luminosities at solar metallicity without
the use of Monte Carlo simulations, and quote some limits for the minimal
initial clusters masses ensuring a relative error lower than 10\% for some
ages and luminosities.  \cite{Brutuc} \cite[see also ][]{BC03}\footnote{A
version of \cite{BC03} synthesis models can be obtained at {\tt
http://www.cida.ve/$\sim$bruzual/bc2003}, or at {\tt
http://www2.iap.fr/users/charlot/bc2003}.}  presents Monte Carlo
simulations in which the stochastic effects on $U-B$, $B-V$, $V-K$, and $K$
for the LMC metallicity are presented as a function of the initial mass of
the cluster. His figures show clearly that there is a bias in the results
of Monte Carlo simulations with respect to the results of analytical
synthesis models; however, this result is not mentioned in the text,
possibly due to the limited space.  Finally, \cite{Gituc} presents Monte
Carlo simulations where the effect of a continuous distribution in the
initial cluster masses is studied.  His results are more appropriate for
the comparison with surveys of real clusters than those that do not
consider a distribution of masses.
 
In all the preceding papers, the evaluation of sampling effects requires
making assumptions on the IMF and the star formation history, a fact that
limits the practical application of the results to real observations.  In
this paper we propose instead a method entirely based on observable
quantities, therefore independent of the IMF and almost independent of the
star formation history, to estimate whether the colors predicted by
synthesis models are biased with respect to real observations, and to
establish when analytical synthesis models cannot be applied and a
statistical formulation is required.  We also describe the relationship of
this method to more sophisticated statistical analyses.  Furthermore, we
suggest examples of possible applications of the method to the analysis of
observational data.  Finally, we discuss the qualities and drawbacks of our
method in comparison to alternative ones.

The structure of the paper is the following: in Sect. 2 we define the
method, and provide a quantitative evaluation of the quantities involved
for a wide range of observational cases; in Sect. 3 we translate the
preceding results in terms of the cluster masses, proposing a mass
criterion to exclude low-statistics clusters from the analysis performed
with synthesis models; in Sect. 4 we show the observational implications of
this work; in Sect. 5 we discuss the limitations of our results; and in
Sect. 6 we draw our conclusions.

\section{The Lowest Luminosity Limit}

One of the most basic limits to the application of evolutionary synthesis
models can be expressed by the following statement:

\begin{center}
{\it The total luminosity of the cluster modeled must be larger than the
individual contribution of any of the stars included in the model}
\end{center}

This obvious statement defines a natural theoretical limit that has not
always been considered when models are applied to real observations.
Whereas in the work by Tinsley \cite[see, e.g.,][]{TG76} it was not
necessary to take this limit into account, due to the observational
limitations at that epoch, the increasing sensitivity of current
instruments has reached a level where this limitation plays a fundamental
role in the interpretation of data.

Defining as the {\it Lowest Luminosity Limit} (hereinafter LLL) the
luminosity of the brightest individual star included in the model, we can
establish a simple luminosity criterion for the application of synthesis
models to the interpretation of observed clusters, imposing the condition
that the cluster modeled be more luminous than the LLL.  While clusters
brighter than this limit may either be well-sampled or not, clusters
fainter than this limit are {\it certainly} misrepresented by synthesis
models.

It is possible to establish a luminosity limit following different
criteria, yielding either weaker or more stringent constraints; but all the
alternative definitions we could think of turned out to either lack
physical meaning, or imply circumstances that do not occur in the modeling
practice.  Nevertheless, the degree of arbitrariness of our definition 
is a hairy problem, and we will discuss it thoroughly in Sect.~\ref{sec:caveats}.

With the definition given above, the LLL is only defined by the isochrone
used and the band under consideration; however, its exact value at a given
age is also weakly dependent on the star formation history.  In the
following we present the values of the LLL computed for a wide range of
parameters.  We have used the isochrones and the integrated magnitudes of
simple stellar population models (hereinafter SSP models, i.e., models that
assume an instantaneous burst of star formation) by
\cite{Gi02}\footnote{Isochrones and simple stellar population results are
available at {\tt http://pleiadi.pd.astro.it/}.}.  We consider seven
different metallicities: $Z$=0.019 (solar), $Z$=0.008, $Z$=0.004,
$Z$=0.001, $Z$=0.0004, $Z$=0.0001, and $Z$=0.0.  The $Z$=0.0 isochrones
correspond to metal-free models by \cite{Ma01}. The other isochrones
correspond to the basic set presented in the web server cited in the
footnote, which combines the results from \cite{Gi00} and \cite{Gi01} for
low- and intermediate-mass stars, with the results by \cite{Be94} and
\cite{Gi96} for high-mass stars, and that includes overshooting and a
simplified Thermal Pulse AGB (TP-AGB) evolution.  Additionally, for
$Z$=0.019, 0.008, and 0.004 we have used the isochrones by \cite{MG01} that
include a more detailed TP-AGB evolution.  Most of the atmosphere models
are taken from ATLAS9 \citep{Cas97}\footnote{NOVER models at {\tt
http://cfaku5.harvard.edu/grids.html}.}.  A more detailed description of
the isochrones, the atmosphere models, and the SSP models can be found in
\cite{Gi02} and in their web server.

To obtain the LLL we have searched for the bolometric luminosity of the
most luminous star in any given isochrone $L^{max}_{bol}(t)$. However,
since the tabulated data also provide the magnitudes at different bands, we
have also obtained the magnitude of the most luminous star in the
Johnson-Cousins-Glass filters $U$, $B$, $V$, $R$, $I$, $J$, $H$, and $K$:
$M^{min}_U(t)$, ..., $M^{min}_K(t)$.

\begin{figure*}[H]
\includegraphics[width=17cm]{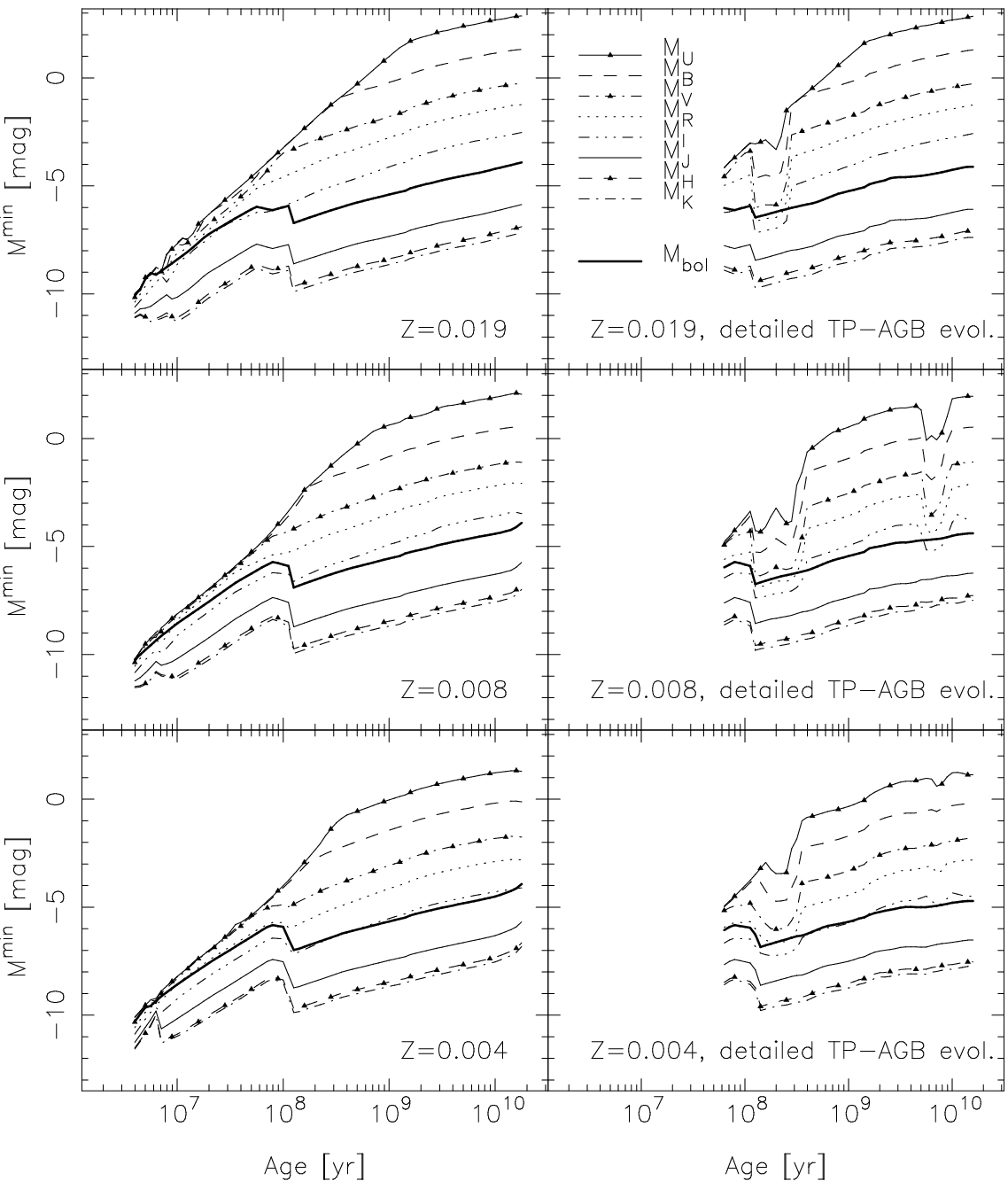}
\caption{LLL in magnitudes for the bolometric luminosity and different
bands, as a function of the age for three different metallicities: top
panels $Z$=0.019 (solar); middle panels $Z$=0.008; and bottom panels
$Z$=0.004. Left panels correspond to isochrones with a simplified TP-AGB
evolution, and right panels to isochrones with a detailed TP-AGB evolution.
The different lines correspond to the following: bold solid line: bolometric magnitude;
solid line with triangles: $U$; dashed line: $B$; dot-dashed line with
triangles: $V$; dotted line: $R$; dash-dot-dot-dotted line: $I$; solid
line: $J$; dashed line with triangles: $H$; and dot-dashed line: $K$.  }
\label{fig1}
\end{figure*}

In general, the most luminous star is also the most evolved, but this
relation does not strictly hold for all the bands nor all the ages. This
fact is well illustrated in Fig. \ref{fig1}, where the LLLs obtained from
isochrones with a simplified TP-AGB treatment (right panels) are compared
with the LLL obtained from isochrones with a detailed TP-AGB evolution
(left panels) for models with $Z$=0.019, 0.008, and 0.004. Whereas
$M^{min}_{U,B,V,R,I}$ depend on the TP-AGB treatment at some ages,
$M^{min}_{bol,J,H,K}$ are almost unaffected by it.

\begin{figure*}[H]
\includegraphics[width=17cm]{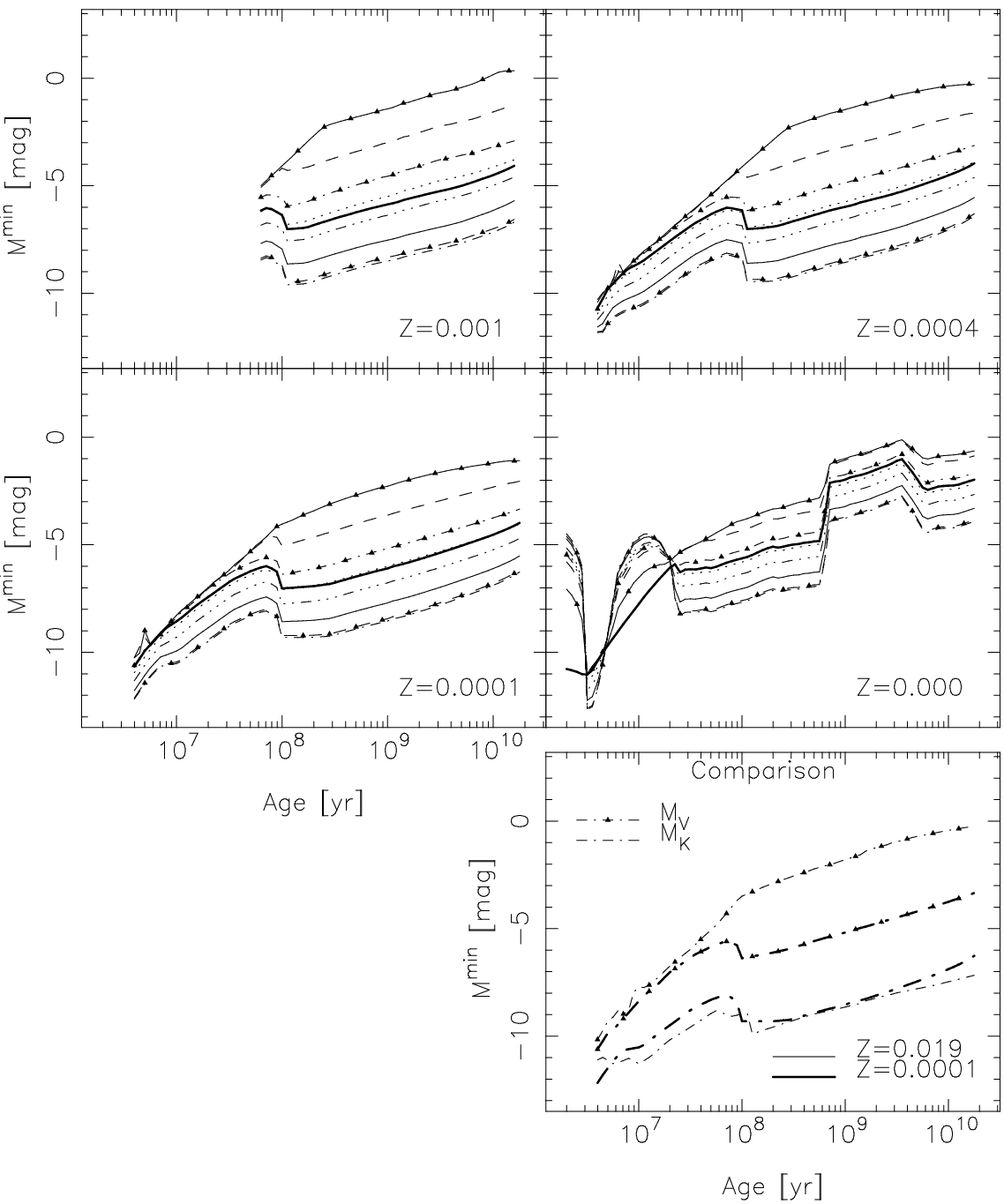}
\caption{LLL in magnitudes for the bolometric luminosity and different
bands, as a function of the age for $Z$=0.001 (top left), $Z$=0.0004 (top
right), $Z$=0.0001 (middle left) and $Z$=0 (middle right). Symbols like in
Fig. \ref{fig1}. The bottom panel compares the LLL for $V$ and $K$ for
$Z$=0.019 and $Z$=0.0001.}
\label{fig2}
\end{figure*}

Figures \ref{fig1} and \ref{fig2} show the values of the LLL in magnitudes
for different ages and metallicities. The figures show that $M^{min}(t)$
evolves with time toward less luminous values for all the bands, and that
blue bands become fainter more quickly than red bands, as expected due to
the cluster evolution: the population becomes redder and fainter as the
cluster evolves.  These two trends combined imply that, as the cluster
ages, the LLL tend to decrease (i.e., become less stringent) in all bands,
doing so more rapidly at blue wavelengths than at red wavelengths.  It is
also interesting to compare the evolution of $M^{min}_{bol}(t)$ with the
various $M^{min}(t)$ at different bands. It can be seen that the larger the
metallicity, the more similar is the evolution of $M^{min}_{bol}(t)$ to the
evolution of $M^{min}(t)$ in red bands, as a consequence of the increasing
fraction of bolometric light that goes into red filters.  The bottom panel
in Fig. \ref{fig2} compares the LLL values for $V$ and $K$ and two extreme
metallicities.  It can be seen that $M^{min}_{K}$ is almost
metallicity-independent.

These figures show the LLL values for the case of single isochrones, that
correspond to SSP models.  However, the LLL can be easily obtained for
different star formation histories.  For example, let us consider a
two-burst system with ages $t_1$ and $t_2$, associated to two minimum
magnitude values $M^{min}(t_1)$ and $M^{min}(t_2)$: then the LLL of this
system is simply the minimum of $M^{min}(t_1)$ and $M^{min}(t_2)$.  In the
case of a cluster with constant star formation history and age $t$, the LLL
at each band is given by the maximum of the luminosity of SSP models in the
age range between $0$ and $t$, which, for the bolometric luminosity, can be
roughly approximated by the ZAMS luminosity of the most massive star
included (this result is not exact because the bolometric luminosity of a
star briefly increases after the ZAMS).  At other bands the maximum
luminosity can be reached much later, thus no easy recipe can be given.  As
a final remark, note that some evolution in the metallicity is expected for
any star formation history different from an instantaneous burst, and this
effect should be included in the modeling in a self-consistent way
\cite[see][and references therein as an example]{FvA}\footnote{The
reference corresponds to the synthesis code {\sc galev} and the model
results are available at {\tt http://alpha.uni-sw.gwdg.de/$\sim$galev/}.}.

Summarizing, the most luminous star in one band is not necessarily the most
luminous star in the rest of the bands, neither is it the star with the
largest bolometric luminosity.  As a consequence, the LLL depends not only
on the age and the metallicity (i.e., the assumed isochrones and atmosphere
libraries), but also on the band considered.  The dependence on age and
metallicity is weaker for the near infrared bands than for the optical
bands.  In particular, the LLL for $K$ is almost metallicity-independent.

\section{Minimal initial cluster masses}

In this section we will translate the concept of LLL into an equivalent
formulation in terms of mass.

Let us recall that evolutionary synthesis models are based on the
convolution of isochrones with the IMF and the star formation history.  For
the case of a SSP, the {\it mean luminosity} in a given band and at a given
age, $l^{ssp}(t)$, results from the sum of the luminosities of individual
stars at the corresponding age as given by the isochrone, $l_i(m_i,t)$,
weighted by the number of stars with initial mass $m_i$ as given by the
IMF, $w_i(m_i)$.  If the sum of the $w_i$ values is normalized, as usual,
to 1 M$_\odot$ transformed into stars from the onset of the burst, the
resulting luminosity will also be normalized\footnote{In the following we
will use the lower case and/or the super-index $ssp$ to refer to normalized
quantities obtained by SSP models, and the upper case for absolute
(denormalized) quantities.}.  The total luminosity of a modeled cluster,
$L^{clus}(t)$, is directly proportional to the initial mass transformed
into stars in the cluster, $\cal M$:

\begin{equation}
L^{clus}(t) = {\cal M} \times {\sum w_i(m_i)\,l_i(m_i,t)} = {\cal M} \times
l^{ssp}(t).
\label{eq:Lclus}
\end{equation}

Then, for a given age and metallicity, we can obtain the total initial mass
transformed into stars from the observed luminosity $L^{clus}$ (or
$M^{clus}$ expressed in magnitudes) and the corresponding normalized value
of $l^{ssp}(t)$ (or $m^{ssp}$):

\begin{eqnarray}
{\cal M} &=& \frac{L^{clus}}{l^{ssp}}, \nonumber \\
2.5 \log {\cal M} &=& m^{ssp} - M^{clus},
\end{eqnarray}

\noindent where we have dropped the explicit reference to $t$ to simplify
the notation.  Now, imposing that $M^{clus}=M^{min}(t)$, we can obtain the
initial cluster mass for each band and age, ${\cal M}^{min}(t)$, for which
the total luminosity of the cluster simulated by a SSP model equals the
luminosity of the most luminous star in the band, $M^{min}(t)$:

\begin{equation}
2.5 \log {\cal M}^{min}(t) = m^{ssp}(t)-M^{min}(t).
\end{equation}

The superindex {\it min} reminds that below this limit the cluster cannot
be modeled by means of a synthesis model.  Note that ${\cal M}^{min}(t)$
depends on the age and the band, but also on the IMF and the star formation
history, since integrated quantities depend on them. Here, we have used the
integrated magnitudes of the SSP models by \cite{Gi02}, which assume the
IMF by \cite{kro01} in its corrected version (his Eq. 6), and a total SSP
initial mass equal to 1 M$_\odot$ in the mass range 0.01 -- 120 M$_\odot$.

\begin{figure*}[H]
\includegraphics[width=17cm]{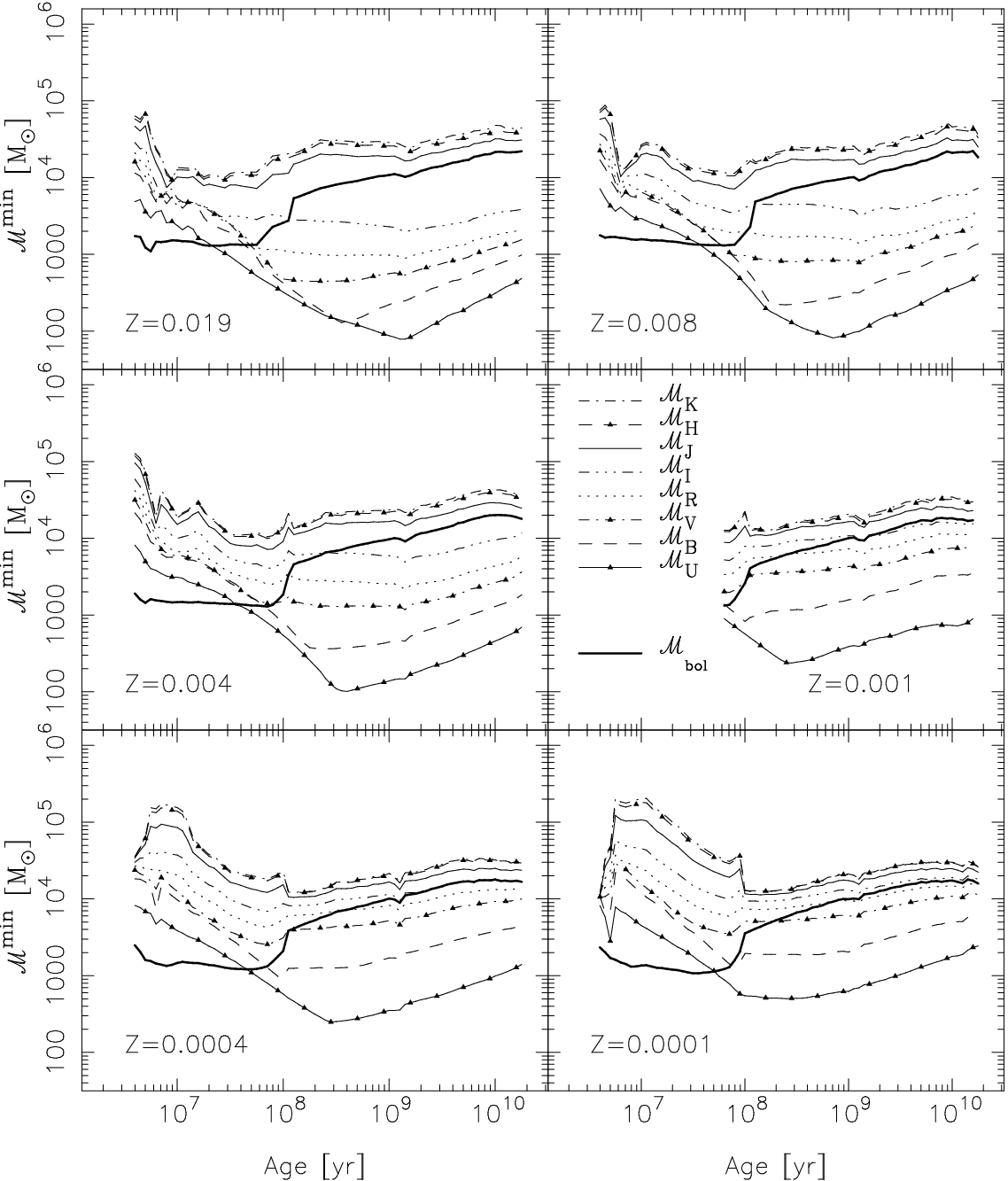}
\caption{${\cal M}^{min}$ for the bolometric luminosity and different
bands, as a function of the age and the metallicity. Symbols as in
Fig. \ref{fig1}.}
\label{fig3}
\end{figure*}

The results for different ages and metallicities are shown in
Fig. \ref{fig3}, which shows that ${\cal M}^{min}_{bol,J,H,K}$ are almost
metallicity independent except during the first stages of the evolution of
the cluster. In the case of optical bands, the lower the metallicity, the
larger ${\cal M}^{min}_{U,B,V,R,I}$. The value of ${\cal M}^{min}$ spans
three orders of magnitude depending on the band, and it takes a value as
large as 10$^4$ M$_\odot$ or even more for the case of near infrared
colors.  A first conclusion we can draw from these figures is that these
values are so high that many observed clusters certainly fall below this
limit, and their properties can {\it by no means} be reproduced by
synthesis models.  This fact is too often overlooked, and we will try to
emphasize it repeatedly throughout the paper.

\begin{figure}
\resizebox{\hsize}{!}{\includegraphics{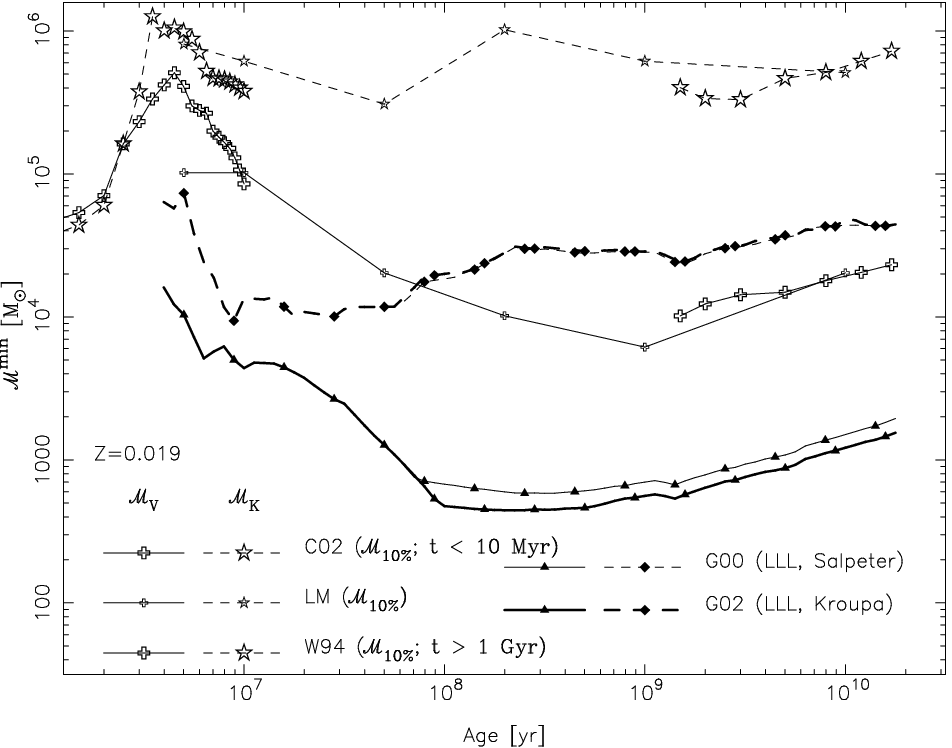}}
\caption{${\cal M}^{min}$ for $V$ (solid line with triangles) and $K$
(dashed line with diamonds) assuming solar metallicity and different IMF and
SSP models: SSP models by \cite{Gi02} with the IMF by \cite{kro01} (G02: bold line), and SSP models by \cite{Gi00} with Salpeter's IMF (G00: narrow
line). The figure also shows the minimal clusters
masses that ensure $\sigma_L/L = 10$\% from \cite{CVGLMH02} models (C02:
big open symbols until age of 10 Myr), \cite{Wor94} models (W94: big open
symbols with ages larger than 1 Gyr) and \cite{LM00} models (LM00: small
open symbols).}
\label{fig4}
\end{figure}

In order to examine the influence of the IMF on this result, we have also
obtained the LLL from the isochrones by \cite{Gi00}, which have been
computed assuming a Salpeter IMF \citep{Sal} in the mass range 0.039 to 100
M$_\odot$ from 63 Myr to 17.8 Gyr.  For a comparison with \cite{Gi02} we
have renormalized these values to the mass range 0.094 -- 120 M$_\odot$ in
such a way that the fraction of mass in the range 1 -- 120 M$_\odot$ for
the Kroupa's and the Salpeter's IMF is the same.  The comparison between
the ${\cal M}^{min}$ values obtained from the two sets of isochrones and
SSP models is shown in Fig. \ref{fig4}.  With this normalization, the
${\cal M}^{min}$ values obtained using the models by \cite{Gi00} coincide
for $V$ with those obtained using the models by \cite{Gi02} at ages smaller
than 80 Myr, and for $K$ in all the age range in common.  The differences
in $V$ for $t>80$ Myr are due to small differences in the LLLs computed
with the different set of isochrones.

\subsection{Minimal initial cluster mass and the estimation of sampling effects
(theoretical point of view)}

Neither the computation of LLL nor that of ${\cal M}^{min}$ allow, by
themselves, an evaluation of the extent of sampling effects in observed
clusters: they just provide an easy-to-use cutoff criterion to discriminate
clusters with severe sampling effects from clusters with more moderate or
negligible sampling effects, without providing a quantitative tool to
estimate their statistical properties.

In spite of this limitation, ${\cal M}^{min}$ is intrinsically related to
sampling effects, and it is worth studying the relation between the
information provided by ${\cal M}^{min}$ and the statistical properties of
clusters.  In such a way, it will be possible to obtain hints on the
necessity to account for sampling effects in the interpretation of observed
data, even without a proper statistical formalism.

To study the relation between the ${\cal M}^{min}$ value and the sampling
effects, we have derived the values of the cluster masses associated to a
relative dispersion of 10\% in the luminosity of a band, ${\cal M}_{10\%}$;
note that a relative dispersion of 10\% in the luminosity means, at zero
order approximation, $\sigma=0.1$ mag in magnitude\footnote{A more detailed
analysis with Taylor expansions to second order shows a small bias of 0.005
mag and an unbiased $\sigma=0.11$. See \cite{CVG02} for more details.}.

To cover a wide age range, we have used the results of three different
synthesis codes.  For young ages, we have used the solar metallicity
results from \cite{CVGLMH02} models\footnote{The models used here do not
include the nebular contribution to the photometric bands. The complete set
of models for 0.1 to 20 Myr, including also the nebular contribution, is
available at {\tt http://www.laeff.esa.es/users/mcs/SED}. }.  These models
have been computed using the Geneva evolutionary tracks with a Salpeter IMF
in the mass range 2 -- 120 M$_\odot$. In these models, sampling effects are
evaluated by means of an {\it effective number of stars} that contribute to
a given observable, ${\cal N}^{ssp}$, which is defined by:

\begin{equation} 
\frac{1}{{\cal N}^{ssp}} =
\frac{\sigma^2({l^{ssp}})}{(l^{ssp})^2}=\frac{\sum w_i\,l^2_i}{(\sum w_i\,l_i)^2},  
\label{eq:neff} 
\end{equation} 

\noindent as first derived by \cite{Buzz89}\footnote{The outputs of
A. Buzzoni synthesis code are available at {\tt
http://www.merate.mi.astro.it/$\sim$eps/home.html}.}. Since the tabulated
${\cal N}^{ssp}$ values are normalized to ${\cal M}$, the value of ${\cal
M}_{10\%}$ is trivially obtained imposing that $\sigma_L/L=0.1$; therefore,
the absolute effective number of stars associated to ${\cal M}_{10\%}$ is
${\cal N}={\cal N}^{ssp}\times {\cal M}_{10\%} =100$.

For intermediate ages, we have used the results quoted by \cite{LM00} that
assume solar metallicity and a Salpeter IMF in the mass range 0.1 -- 120
M$_\odot$.  The models have been computed with the population synthesis
code {\sc pegase}\footnote{Available at {\tt
http://www2.iap.fr/users/fioc/PEGASE.html}.} \citep{FRV}.

For old stellar populations ($t> 1.5$ Gyr) we have used the solar
metallicity SSP models by \cite{Wor94}\footnote{Available at {\tt
http://astro.wsu.edu/worthey/}.} computed for a Salpeter IMF in the mass
range 0.1 -- 2 M$_\odot$ and normalized to ${\cal M}=10^6$ M$_\odot$. This
author does not compute directly sampling effects, but they can be inferred
from the fluctuation luminosities, $\bar{l}^{ssp}$. These $\bar{l}^{ssp}$
(or $\bar{m}^{ssp}$ expressed in magnitudes) are defined as:

\begin{equation}
\bar{l}^{ssp}=\frac{\sum w_i\,l^2_i}{\sum w_i\,l_i},  
\label{eq:barm} 
\end{equation} 

\noindent and they are computed for the evaluation of surface brightness
fluctuations.  From Eqs. \ref{eq:neff} and \ref{eq:barm} it is found that
${\cal N}^{ssp}\times \bar{l}^{ssp} = l^{ssp}$ \cite[see][for a general
description of the relation between ${\cal N}^{ssp}$ and the brightness
fluctuations]{Buzz93}.  The corresponding ${\cal N}^{ssp}$ can be obtained
as a function of $\bar{l}^{ssp}$ and $l^{ssp}$ (or $\bar{m}^{ssp}$ and
$m^{ssp}$ in magnitudes) using the following formulae:

\begin{eqnarray}
{\cal N}^{ssp} \times 10^6 &=& \frac{l^{ssp}}{\bar{l}^{ssp}}, \nonumber\\
2.5 \log {\cal N} + 15 &=& \bar{m}^{ssp} - m^{ssp},
\label{wor}
\end{eqnarray}

\noindent where the factor $10^6$ (or $15=2.5 \log 10^6$ in magnitudes), is
used to renormalize his tabulated data to ${\cal M}=1$ M$_\odot$ following
a Salpeter IMF slope in the mass range 0.1 -- 2 M$_\odot$.

Note that \cite{LM00} and \cite{CVGLMH02} quote the dispersion for the
monochromatic luminosities at the effective wavelength of the band
\citep[see][for the definition of effective wavelength]{K52} whereas
\cite{Wor94} (Eq. \ref{wor}) gives the dispersion of the integrated
luminosity of the band.  In spite of this difference, these results can be
directly compared: in fact, wide-band luminosities can be estimated, with
an accuracy of 2 -- 3\%, by multiplying the monochromatic luminosities at
the effective wavelength by a constant value, which represents the absolute
flux density in the band \citep{K52,J66}; since this transformation only
implies the multiplication by a constant, which cancels out in the
computation of the variance, the relative dispersion of the monochromatic
luminosity is equal to the relative dispersion of the luminosities in the
band obtained from the exact integration of the spectrum over the filter
response.  Therefore, the results by \cite{LM00} and \cite{CVGLMH02} and
those by \cite{Wor94} quoted above can be directly compared.

In all the cases we have renormalized the resulting values to a Salpeter
IMF in the mass range 0.094 -- 120 M$_\odot$. The results are shown with
open symbols in Fig. \ref{fig4}. A first comparison among the ${\cal
M}_{10\%}$ results show that they are quite consistent with each other,
with some differences that can be attributed to the difference between the
${\cal N}$ formalism and the method applied by \cite{LM00}.  The figure
also shows that the evolution of ${\cal M}^{min}$ is quite similar to the
evolution of ${\cal M}_{10\%}$. The differences in the two curves range
between 0.98 and 1.5 dex (i.e., factors between 8 and 30 in initial cluster
masses) depending on the age and the band.  Therefore, for the assumed IMF
and rounding up numbers, we can say that ${\cal M}^{min}$ is always at
least a factor 10 below ${\cal M}_{10\%}$; this implies an ${\cal N}$ value
lower than 10 and relative dispersions larger than 32\% ($\sigma > 0.35$
mag). For these values of ${\cal N}$, Poisson statistics produce non
negligible probabilities of zero effective sources, and hence the presence
of biases in colors, as shown in \cite{CVG02}.

The relation between ${\cal N}$ and the occurrence of dispersion and of a
bias can be easily understood in the following terms.  Let us take as an
example the case of $V-K$, and assume that $L^{LLL}_K$ corresponds to the
luminosity of a star in the Red Supergiant (RSG) phase. Let us also assume
the case of a 10 Myr old burst with solar metallicity, where, according to
analytical models, more than 90\% of the luminosity in $K$ is due to
RSGs. For this case, comparing the corresponding ${\cal N}^{ssp}(L_K)$
value with the (normalized) number of RSGs, $n^{ssp}_{RSG}$, it is found
that $n^{ssp}_{RSG} \sim 0.9 \times {\cal N}^{ssp}(L_K)$ \cite[see][and the
web server mentioned above]{CVGLMH02}. For simplicity we will assume that
${\cal N}(L_K) = N_{RSG}$, the absolute number of RSGs. Finally let us
assume that ${\cal N}$ follows Poisson statistics, or, in terms of RSGs,
that the number of such stars in different clusters is distributed
following a Poisson distribution with a mean value $N_{RSG}$.

Since the contribution of RSG stars has a small influence on $V$ and a
large influence on $K$, clusters in a mass range in which variations of
$\pm 1$ in the number of such stars are relevant will have colors
considerably redder (dominated by an excess of RSGs) or bluer (due to a
deficit of RSGs) than those predicted by analytical synthesis models.
Furthermore, if the mass of the cluster is such that $1 < {\cal N} < 10$,
according to Poisson statistics there is a fair probability that the
cluster has no RSGs at all. In this last case $(V-K)^{clus}$ will be more
similar to the colors of main sequence (MS) stars than to the colors of SSP
models (i.e., there will be an excess of blue clusters in a survey of
clusters with this mass, age, and metallicity). For these values of ${\cal
N}$ the dispersion in the colors will be the largest. Finally, if ${\cal
N}$ takes values lower than 1, there will be an important fraction (or, in
extreme cases, even a majority) of clusters without RSG stars, and then,
the mean value of the observed color $(V-K)^{clus}$ will be biased with
respect to the resulting color $(V-K)^{ssp}$ of a synthesis model. On the
other hand, the dispersion will decrease since the range of possible
$(V-K)$ values will be smaller with respect to the case of larger ${\cal
N}$ values.

The situation for the case of ${\cal N}=5.5$ is illustrated in
Fig. \ref{fig5}, which shows the results of 10$^4$ Monte Carlo simulations
for clusters with 10$^3$ stars in the mass range 2--120 M$_\odot$.  The
analytical value of $(V-K)^{clus}$ is shown by a vertical dashed line, and
its distribution in the set of simulated clusters is shown by the bold
solid line.  Note that the mean cluster mass of these simulations is ${\cal
M}=2\times 10^4$ M$_\odot$, a value larger than the minimal mass value
${\cal M}^{min}=1.5\times 10^4$ M$_\odot$: this fact, and the considerable
width of the $(V-K)^{clus}$ distribution, confirms that a considerable
dispersion in the observables is still expected for cluster masses larger
than ${\cal M}^{min}$, up to values as large as $10\times {\cal M}^{min}$
(Figure~\ref{fig4}).  According to Poisson statistics, there is a 0.4\%
probability of finding a cluster without any RSGs.  Indeed, there is a
small accumulation of simulations with $(V-K)^{clus}$ values around $-0.6$
mag, and the number of such clusters is about 40 (i.e., 0.4\% of the
total). Note also that the distribution is negatively skewed, i.e., it
tends to cut off sharply in the red and extend toward the blue. This means
that either there is a deficit of RSGs, or that the RSG stars in the
cluster have luminosities lower than the one that defines the LLL.

This statistical interpretation depends in fact only on the ${\cal N}$
value, independently of it being related to a physical number of stars
\cite[see the figures presented in][as an example]{Brutuc}. In general, the
distribution of colors will be skewed toward the band with larger ${\cal
N}^{ssp}$, i.e., toward the blue in the present example.

Note that this interpretation has been done in terms of cluster masses (or,
equivalently, ${\cal N}$): the mass and the age of the cluster are fixed,
and the dispersion in the luminosities is produced by the random
differences in the stellar mass spectrum.  However, the cluster mass is not
an observable.  So a different approach is needed to deal with the
observational problem, and it is more useful to use directly the LLL, as we
show in the next section, where the remaining features of Figure~\ref{fig5}
will be discussed.

\section{Applications of the  Lowest Luminosity Limit (observational point
of view)}

Up to this point, we have discussed the statistical properties of real
clusters, trying to answer the following theoretical question: {\it What is
the statistical dispersion to be expected in the observables of clusters
with given mass (or number of stars) and age?}  Although answering this
question surely provides a deep insight in the analysis of stellar
clusters, it must be kept in mind that the observational approach to this
problem differs from the theoretical vision: a simple reason is that, when
observations are made, neither ${\cal M}$ nor the age of the observed
clusters are known.  The observational question can be instead put in these
terms: {\it Given an observed value of the luminosity, which are the
distributions of ${\cal M}$, age, and metallicity consistent with the
observations?}

Unfortunately, as we have repeatedly stated earlier in the paper, the LLL
method is not a sophisticated tool, and its reach is very limited.  More in
general, the last question cannot be addressed with the current theory
available, and a more elaborated theoretical study of this subject is
needed; to this respect we want to remind again the work from \cite{Gituc}
as the most plausible direction to which the theoretical evaluation of the
dispersion must be focused.  However, the concept of LLL is powerful enough
to allow us some applications to real observational problems.  Since one of
the constraints imposed by observations is the existence of a luminosity
limit, in this section we will explore the consequences of having a cutoff
in luminosity in surveys of star clusters.

\begin{figure}
\resizebox{\hsize}{!}{\includegraphics{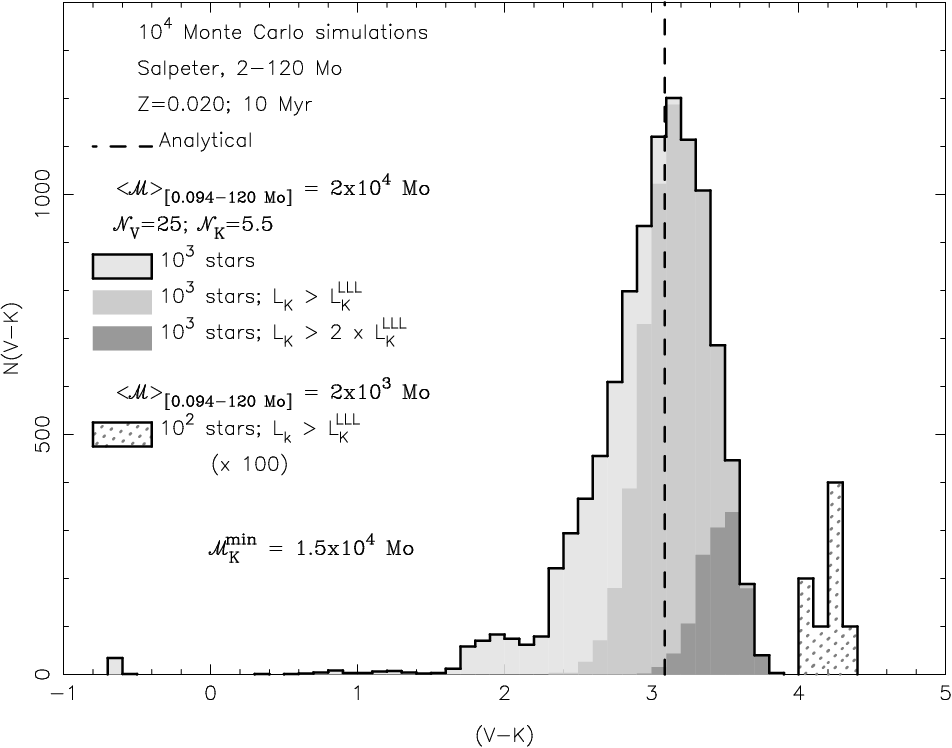}}
\caption{Probability density distribution for $V-K$ obtained from 10$^4$
Monte Carlo simulations of 5.5 Myr, with 10$^3$ stars in the mass range
2--120 M$_\odot$ (which corresponds to $\cal M$=2$\times 10^4$ in the mass
range 0.094--120 M$_\odot$). The $V-K$ distribution for different 
cutoffs in $L_K$ is shown by different shadings (see text). 
The distribution for clusters with 10$^2$ stars
$L_K >  L_K^{LLL}$ is also shown.}
\label{fig5}
\end{figure}

In Figure~\ref{fig5} we described the $V-K$ distribution of a sample of
model clusters with fixed number of stars, and noted that such distribution
is negatively skewed.  Note that a fixed number of stars correspond roughly
to a fixed cluster mass.  Now, let us consider only the subset of clusters
with a luminosity larger than the LLL in $K$, $L_K^{LLL}$. This subset is
shown by the light-shaded histogram, and it can be seen that its mean color
is somewhat redder than the predictions of SSP models.  The key point here
is that {\it this behavior indicates that the statistics of the clusters is
low}: in fact, at $t=10$ Myr an important fraction of the luminosity in $V$
is provided by MS stars, whereas the luminosity in $K$ is completely
dominated by RSGs, which are intrinsically scarcer than MS stars.
Therefore the spread in $K$ among the clusters of the sample is
comparatively large, and a cutoff in luminosity, leaving out the faintest
clusters, sensibly alters the mean $K$ value of the remaining subset.
Schematically, we can say that in $L_K$ limited samples the mean $K$
luminosity increases and the $V$ luminosity is barely affected, thus $V-K$
increases.  This is confirmed by the application of the more restrictive
constraint $L_K^{clus} > 2 \times L_K^{LLL}$, which is shown by the
dark-shaded histogram: the number of clusters that fulfill this constraint
is even lower, and they are redder than the predictions of SSP models.

So far for clusters with (roughly) the same mass.  Now, let us consider the
behavior of simulated clusters with different masses.  To this aim, we will
assume in the following that the distribution of cluster masses follows the
law $\psi({\cal M})\propto {\cal M}^{-2}$, as proposed by \cite{ZF99}.
Whether this mass distribution correctly represents real clusters does not
concern us for the moment: for the sake of the argument, it is enough to
assume just any law. In fact, at the end of this section we will mention a
possible caveat of this particular law, which might possibly imply a bias
in the method used to derive this law.  We have performed 10$^4$ Monte
Carlo simulations of clusters with 10$^2$ stars (${\cal M}=2\times 10^3$
M$_\odot$), which have a mean mass roughly 1/10 that of the simulated
clusters discussed up to now.  Given the mass distribution law assumed, we
have multiplied each bin by 100, to reproduce the expected number of
clusters with this mass compared with clusters with mass ten times larger.

The dotted histogram in Figure~\ref{fig5} shows the distribution of those
of such clusters that also fulfill the condition $L_K^{clus} > L_K^{LLL}$.
Note that the histogram peaks around $(V-K)^{clus}=4.2$, i.e. these
clusters have extremely red colors; these behavior is consistent with the
interpretation of being an effect of sampling, which in clusters of 10$^2$
stars is much more severe than in clusters of 10$^3$ stars.

This argument can be repeated for cluster samples of any mass value. If we
extrapolate the result to a continuous cluster mass distribution, we
readily realize that the resulting histogram will have a cutoff in the blue
(small $V-K$ values), and a long tail in the red (large $V-K$ values):
i.e., it will be {\it positively skewed}, contrarily to the histogram of a
complete (non-luminosity limited) sample of clusters with a fixed mass.
Generalizing to different colors, the observed distribution of colors in a
luminosity-limited sample will be skewed toward the band with the lowest
${\cal N}^{ssp}$ value.

Two main conclusions can be drawn from these results.  First, the shape of
the color distribution of a luminosity-limited sample of clusters may be
used to constrain the underlying cluster mass distribution, if other
cluster parameters, such as age and metallicity, are known.  To this
respect note the particular shape emerging from an extrapolation of our
example to more mass values is just a consequence of having assumed a
particular mass distribution law: in an observed sample, the shape may in
principle be different.  Second, these results also show that the LLL is an
useful (probably, too conservative) criterion to detect the clusters with
severe sampling effects.  Indeed, in our examples the color of the subsets
with $L_K^{clus} > L_K^{LLL}$ is different from that of the complete set,
indicating that the sample suffers from sampling effects: had the mean mass
of the clusters in the sample been sufficiently larger, no clusters would
have been excluded by the luminosity criterion.  Note also that the fact
that the application of the luminosity cutoff changes the mean color of the
sample implies that {\it all} the clusters of the sample suffer from
incomplete sampling, and not only those that are excluded: that is, having
a luminosity larger than the LLL is a condition necessary but not
sufficient for the meaningful application of synthesis models;
Equivalently, the LLL criterion detects some, but not all, of the clusters
with poor statistics.

To conclude this section, a note of caution about the determination of the
mass distribution law by \cite{ZF99}: these authors derived their law by
considering only clusters with $M_V$ brighter than -9 to avoid
contamination of the sample by individual stars, hence applying a selection
criterion that is more or less the observational counterpart of the LLL
method.  However, the $M_V^{min}$ we obtain here for the age range they
consider (2.5 $< t <$ 6.3 Myr) lies between -10 mag and -9 mag: therefore,
their analysis could possibly be affected by the sampling effects and the
bias we are discussing.

\subsection{The distribution of Globular Clusters}

A clear example of a skewed distribution is given by the sample of Globular
Clusters (GCs) by \citep{GKP99}, a fact suggesting that the properties of
these objects might be affected by sampling effects.  This statement might
seem surprising, for GCs are the paradigm of well-populated objects; stars
in GCs occupy the most populated part of the IMF, so one would naively
conclude that they cannot be possibly affected by sampling effects.
However, the importance of sampling effects does not only depend on the
absolute number of stars in a given mass range, but also on the
evolutionary time scale considered: indeed, globular clusters also contain
bright stars in low-populated evolutionary phases, such as the RGB and AGB
phases.  Hence sampling effects may also show up at old ages.

Let us illustrate the problem by the application of the LLL to GCs in NGC
5128. \cite{Rej01} presents detailed photometry of GCs in NGC 5128. We have
used the result from this author because her plots can be easily reproduced
with the data and the indications given in the paper.  She compares the
position of the GCs in color-color diagrams with the results of synthesis
models, and notices that the clusters lie slightly to the right and below
the model lines in the $V-K$ vs. $U-V$ plane.  She mentions two possible
explanations: (i) a difference between observations and SSP models,
particularly significant in $U-V$, which has been hypothesized by \cite{BH}
as a consequence of problems in the atmosphere libraries used by synthesis
models \cite[see also][]{BK78}; (ii) an additional offset may arise from
the difference between the Bessell and the Johnson $U$-band transmission
curves. These facts might explain the offsets between SSP models and the
mean color of the observations, but they do not explain neither the
observed dispersion nor the shape of the distribution.  Let us now study
these discrepancies when the LLL is taken into account.

\begin{figure}
\resizebox{\hsize}{!}{\includegraphics{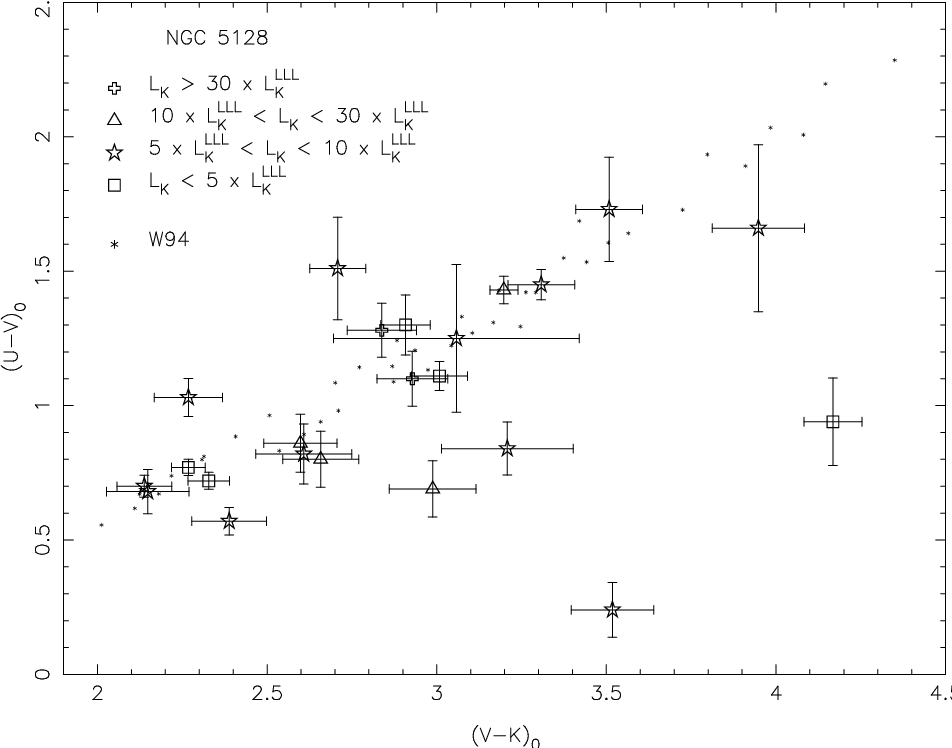}}
\caption{Color-color diagram for Globular Clusters in NGC 1528. Crosses
correspond to clusters with $L^{clus}_K/L^{LLL}_{K} > 30$
($t$= 1 Gyr); triangles to $L^{clus}_K/L^{LLL}_{K} \in [10,30]$; stars to 
$L^{clus}_K/L^{LLL}_{K} \in [5,10]$; 
and squares to $L^{clus}_K/L^{LLL}_{K} < 5  $. Small
asterisks corresponds to \cite{Wor94} models for all the ages and
metallicities quoted by the author.}
\label{fig6}
\end{figure}

Following \cite{Rej01}, we have assumed a distance modulus of 27.8 and we
have corrected for extinction the observed photometric bands with a mean
$E(B-V)=0.1$ using $A_U=4.40$, $A_V=3.1$, $A_K=0.38$, according to the
values quoted in the ADPS project \citep{MM00}\footnote{Available at {\tt
http://ulisse.pd.astro.it/Astro/ADPS/}.}.  We have grouped the GCs
according to their $L_K^{clus}$ luminosities: $L^{clus}_K/L^{LLL}_{K} < 5
$, $L^{clus}_K/L^{LLL}_{K} \in [5,10]$, $L^{clus}_K/L^{LLL}_{K} \in
[10,30]$, and $L^{clus}_K/L^{LLL}_{K} > 30$.  We have used the $L_K^{LLL}$
value at 1 Gyr, that corresponds to $M_K^{min}=-8.5$ mag.  The resulting
color-color diagram is shown in Fig. \ref{fig6}.  The results of SSP models
from \cite{Wor94} have also been plotted for comparison. Note that a
rigorous comparison should include the confidence intervals for the results
of SSP models; however, this would require knowledge of the correlation
coefficients between the different luminosities, which, unfortunately, have
only been evaluated for the case of young stellar populations
\cite[see][]{Clan,CVGLMH02}.

The figure shows several interesting results.  First of all, the cluster
sample tends to have $V-K$ colors redder than the predictions of SSP
models, as is expected from a luminosity limited survey.  Second, there is
a statistical trend of the position in the plane with the $L_K$ value: (i)
The first bin of clusters, the one with $L^{clus}_K/L^{LLL}_{K}< 5$,
contains five of the 23 clusters plotted, and one of them is the cluster
that disagree the most with the SSP results.  (ii) Of the twelve clusters
with $L^{clus}_K/L^{LLL}_{K} \in [5,10]$, five disagree with the results of
synthesis models.  (iii) There are four clusters with
$L^{clus}_K/L^{LLL}_{K} \in [10,30]$, and only one of them falls far from
the results of SSP models.  (iv) The two clusters with
$L^{clus}_K/L^{LLL}_{K} > 30$ are reproduced reasonably well by SSP models.
In summary, clusters with lower $L_K$ tend to deviate more from SSP models
than luminous clusters, as can be expected when sampling effects are
present.

\section{Caveats about self-consistency and limitations}\label{sec:caveats}

In the present work we have established a luminosity limit, the LLL, for
the application of synthesis models to observed clusters.  The definition
we chose for the LLL might raise objections, in at least two different
ways.  First, the LLL is implicitly defined by the isochrones and the
atmosphere libraries assumed in the code; a natural doubt is therefore
whether assuming different input libraries in the code would change our
results.  Second, there are evolutionary phases that are characterized by
extremely high luminosities and short lifetimes.  According to our
definition, it is these phases that determine the value of the LLL, at the
ages they occur; nevertheless, they have a very small probability of being
observed in real clusters. Therefore we face the apparent oddity of setting
a limit much more restrictive than necessary.

This section is devoted to clarify these points and to convince the reader
that our definition of the LLL is operationally good and, to a satisfactory
extent, physically based.  We will also discuss its residual degree of
arbitrariness, and the limitations it implies.  Finally, we will also
explain why, by virtue of its limited reach, the exact definition of the
LLL is not very crucial in practice.

\subsection{Dependence on input isochrones and model atmospheres}

A possible objection to our definition of the LLL is that it depends on the
input libraries (isochrones and atmosphere models) assumed by the synthesis
code, suggesting that it is not physically grounded.  For example, the LLL
can be lowered if some stages of stellar evolution are not included in the
computations of the synthesis code, as is often the case for the Asymptotic
Giant Branch (AGB) phase, which is neglected, e.g., in starburst-oriented
codes such as the code by \cite{CMH94} and {\sc
starburst99}\footnote{Available at {\tt
http://www.stsci.edu/science/starburst99/}.}  \citep{SB99}.  Obviously, the
results of evolutionary synthesis models become intrinsically incomplete at
the ages where the evolutionary stages neglected are relevant; in the case
of the AGB phase, ages older than 50 Myr.  Therefore, under these
circumstances the concept of LLL cannot be applied, not because it is
ill-defined, but rather because the code should not be used in the first
place.

Furthermore, the agreement shown in Fig.~\ref{fig4} between the trend of
our predictions for ${\cal M}^{min}$ and the ${\cal M}_{10\%}$ values by
various authors, which have been obtained with different isochrones and
atmosphere libraries, suggests that the LLL value does not substantially
depend on the input libraries assumed - provided they include the same
evolutionary stages. This conclusion should not surprise too much, since
the results of different theoretical models in the fields of stellar
evolution and atmosphere modeling are substantially converging.  It must be
noted, however, that it should not be of great concern even if this were
not the case, since at the LLL level one must only worry about
self-consistency, i.e. that the LLL is taken from the same input libraries
that enter the synthesis code; the judgment on the input data quality has
already been made with the choice of the synthesis code.  Stated otherwise,
the choice of a synthesis model is logically prior to the computation of
the LLL, and it already implies trusting the input data the model is based
on.

\subsection{Dependence on short, very luminous evolutionary
phases}

Some readers may have noted that our definition does not involve stellar
lifetimes, but just their luminosities.  Such a definition might seem too
restrictive in those evolutionary phases characterized by very luminous but
short-lived objects: Supernovae (SNe), luminosity peaks of Luminous Blue
Variable stars, and recurrent He-shell flashes in TP-AGB stars are cases in
point.  Due to their short duration, the probability of observing such
phases in real clusters is vanishingly small, and they do not usually
contribute to the observed luminosities.  Yet, from a theoretical
standpoint these phases provide the maximum luminosity: hence they
determine the LLL value at the corresponding ages.  Therefore, it may seem
that our definition is at risk of giving too much weight to luminous,
short-lived evolutionary phases; if this were the case, the definition
would be too restrictive, and it would mistakenly ban the application of
synthesis models to cases where they would otherwise be applicable. Although
correct in principle, this conclusion does not hold in practice, as we will
make clear in the following.

First, very short evolutionary phases are not usually included in
isochrones, due to the same reason for which they should not determine the
LLL: they are not expected to be observed.  Of course, this argument would
no longer be valid should these rapid phases be included in isochrones: but
users of synthesis models should always take the elementary caution of
documenting themselves about the properties, assumptions, and limitations
of the models they use.

Second, it is not easy finding cutoff criteria alternative to the LLL and
equally easy to apply, to discriminate systems with poor statistics from
systems with good or unknown statistics.  An alternative definition could
be based, for example, on the comparison of the cluster luminosity with the
{\it sum} of the luminosities of all the individual stars included in the
isochrone (rather than the most luminous one alone, as in our definition).
Unfortunately, such criterion depends on the mass resolution - the number
of points - of the isochrone, which in turn depends on the individual,
idiosyncratic interpolation philosophy of each code. Therefore, such
definition would yield profoundly different results for different codes,
even if the same set of isochrones and atmosphere libraries were adopted.
Still another definition could rely on the {\sl energy} emitted by a star
during a particular phase, rather than its luminosity, in order to avoid
the problem that arises with very luminous but short stellar phases.
However, it can be seen that in this case the limit depends on how
evolutionary phases are defined.  For example, two possible alternatives
would be either (i) ``standard'' evolutionary phases (e.g., Wolf-Rayet,
RSG, AGB stars, and so on), or (ii) the interval around each point on the
isochrone.  In either case, the numerical result would depend on the
adopted definition of evolutionary phase, hence on an arbitrary choice.

As a final remark, it should be emphasized that neither the LLL nor ${\cal
M}^{min}$ define the statistics and the confidence levels of synthesis
models, which are defined, for example, by the ${\cal N}$ formalism.  In
this sense, the LLL is just a two-state switch that indicates whether in
each given case a statistical approach is mandatory; if it is not, it is
still possible that sampling effects do play a role.  In epidemiological
terms, the LLL criterion is extremely specific, but not very sensitive: it
does not give false positives (``good'' clusters with rich statistics
misunderstood for ``bad'' clusters with poor statistics), but it might give
many false negatives (``bad'' clusters with poor statistics not
identified). As such, it gives a much more basic information than ${\cal
N}$, hence it is not too worrisome whether its definition has a certain
degree of arbitrariness.  Ideally, the predictions performed by any
synthesis code should be accompanied by a statistical analysis of the
system under study, bypassing the need to resort to coarse indicators like
the LLL.

\section{Summary and conclusions}

In this paper, we have drawn the attention to a basic condition that must
be fulfilled in order to interpret observed stellar clusters by means of
synthesis models:

\begin{center}
{\it The total luminosity of the cluster modeled must be larger than the
individual contribution of any of the stars included in the model}.
\end{center}

In particular, the luminosity of the most luminous star included in the
model defines a Lowest Luminosity Limit below which a cluster suffers from
severe sampling effects, and it cannot be modeled by an analytical
synthesis code.  We have emphasized that, although our method does not
recognize {\it all} the clusters affected by sampling effects, by virtue of
its simplicity it can be used to separate out the most critical cases from
large surveys without the need to perform a sophisticated and
time-consuming statistical analysis.  Transient events must be considered
in the LLL only if they are included in the results of SSP computations; if
they are not included, the definition of the LLL is fully meaningful, under
the assumption that such events are not present in the observations.
Although this assumption might seem, at first glance, a very restrictive
one, it does not in fact limit the application of synthesis models, since
the SSP results themselves are also only valid under the same assumption.

We have obtained the Lowest Luminosity Limit for a wide range of ages and
metallicities.  In turn, the Lowest Luminosity Limit has been used to
obtain the associated mass limit ${\cal M}^{min}$, which is the cluster
mass below which the colors may be severely biased with respect to the
results of synthesis models.  The limit depends on the age and the
metallicity of the cluster and is more stringent in the near infrared
bands; in particular, it has a value between 10$^4$ and 10$^5$ M$_\odot$
for $K$.  We have also shown that the luminosities in different bands of
clusters with mass ten times larger than ${\cal M}^{min}$ have relative
dispersions of 10\% at least: that is, ${\cal M}^{min}$ can be used to
estimate a lower limit to the expected dispersion of luminosities

As observational applications, we have shown that in surveys that reach
luminosities near the Lowest Luminosity Limit the color distributions will
be skewed toward the luminosity with the lowest ${\cal N}^{ssp}$ value, and
that the skewness may be used to constrain the distribution of initial
cluster masses ${\cal M}$, provided the other relevant cluster parameters
are known.  We have also analyzed a sample of Globular Clusters in NGC
5128, showing that sampling effects are probably playing a role, and that
they can explain, at least partially, the bias of the observed colors with
respect to the predictions of synthesis models.

Finally, we have discussed the assumptions underlying our definition of the
LLL, its virtues, and its drawbacks.  In particular, on the positive side
we have stressed the extreme simplicity of the LLL as a criterion to
recognize in an economical way the occurrence of poor statistics in observed
clusters; on the negative side, we have highlighted the limited capability
of the LLL criterion to recognize {\it all} the cases of poor statistics.

The preceding results have been obtained from very basic concepts implicit
in the modeling performed by synthesis codes.  In particular, the
definition of the LLL is very simple, and its computation can be as
elementary as finding a maximum in a table, at least in the case of SSP
models.  But this simplicity may be deceiving, for simple as the concept
may be, it has far-reaching consequences, which have been too often
overlooked in the analysis of stellar clusters.  Furthermore, although the
simplicity of the concept presented here might disappoint those who are
fond of spectacular or complicated results, we consider it as an asset
rather than a drawback, since the simplicity of our method may serve as an
incentive to those who realize the importance of sampling effects but do
not know how to deal with them.  Therefore, our aim in the present work is
twofold and goes beyond the particular results we have described in this
paper: in the short term, we want to provide a very simple sieve to detect
rapidly clusters with poor statistics; in the long term, we hope to
motivate other researchers, both theoreticians and observers, to move in
this direction and explore the implications of sampling effects in the
analysis of the integrated light from clusters.  In particular, we urge all
the model makers to include a proper statistical treatment in their code,
and all the model users to take into account the effects of sampling in the
interpretation of the data.  Current synthesis models are an optimum tool
for the interpretation of the average properties of stellar systems, but
they need to be updated to keep pace with the new observational data.

\begin{acknowledgements}
We thank the referee, Leo Girardi, for helping us to improve this paper
through his constructive criticism and invaluable suggestions.  We also
want to acknowledge Marina Rejkuba for very useful comments during the
preparation of the manuscript.  Useful discussions have been provided by
Alberto Buzzoni, David Valls-Gabaud, Enrique P\'erez, Ariane Lan{\c{c}}on
and Jo\~ao Francisco C. Santos Jr.  We also want to acknowledge Grazyna
Stasi{\'n}ska, Rafael Guzm\'an, and Marina Rejkuba for pushing us to
publish this paper.  MC has been partially supported by the AYA 3939-C03-01
program.  VL is supported by a Marie Curie Fellowship of the European
Community program {\it Improving Human Research Potential and the
Socio-Economic Knowledge Base} under contract number HPMF-CT-2000-00949.
\end{acknowledgements}

\bibliographystyle{apj}


\end{document}